\documentclass{hpc-ua}

\begin{document}

\usefont{T2A}{ftm}{m}{n}

\selectlanguage{english}
\selectlanguage{russian}

\title{Complex Workflow Management and Integration of Distributed Computing Resources by Science Gateway Portal for Molecular Dynamics Simulations in Materials Science}
\author{Yuri Gordienko, Lev Bekenov, Olexandr Gatsenko, Elena Zasimchuk, Valentin Tatarenko}
\affiliation{G.V.Kurdyumov Institute for Metal Physics, National Academy of Sciences of Ukraine, Kyiv, Ukraine}
  \email{gord@imp.kiev.ua}

\maketitle

\begin{abstract}
The ``IMP Science Gateway Portal'' (http://scigate.imp.kiev.ua) for complex workflow management and integration of distributed computing resources (like clusters, service grids, desktop grids, clouds) is presented. It is created on the basis of WS-PGRADE and gUSE technologies, where WS-PGRADE is designed for science workflow operation and gUSE - for smooth integration of available resources for parallel and distributed computing in various heterogeneous distributed computing infrastructures (DCI). The typical scientific workflow with possible scenarios of its preparation and usage is considered. Several typical science applications (scientific workflows) are considered for molecular dynamics (MD) simulations of complex behavior of various nanostructures (nanoindentation of graphene layers, defect system relaxation in metal nanocrystals, thermal stability of boron nitride nanotubes, etc.). The advantages and drawbacks of the solution are shortly analyzed in the context of its practical applications for MD simulations in materials science, physics and nanotechnologies with available heterogeneous DCIs.
\end{abstract}


\section{Introduction}

During the last decades new nanoscale materials (nanomaterials) became of a great interest, because of their unique nanoscale structure and properties, for example, nanocrystals, graphene, carbon nanotubes, boron nitride nanotubes (BNNT), and their complexes. Molecular dynamics (MD) simulations of nanoscale structures, processes, and properties are very promising. They are especially perspective in the wide range of physical parameters, because of the possibility for "parameter sweeping" parallelism. Actually, MD simulations can be performed in a brute force manner in the available heterogeneous distributed computing infrastructure (DCI) based for example on desktop grids, service grids, clusters, cloud resources. The recent advances in computing algorithms and infrastructures, especially in development of DCIs, allow us to solve these tasks efficiently without expensive scaling-up. DCIs on the basis of the XtremWeb-HEP \cite{xtremweb2005}, OurGrid \cite{ourgrid2006}, BOINC SZDG \cite{szdg2006}, and EDGeS \cite{edges2008} platforms for high-performance distributed computing are very promising way to use the heterogeneous computing resources, especially by means of the science gateway (SG) technology on the basis of WS-PGRADE platform for workflow management and gUSE technology for integration of DCI-resources \cite{wspgrade2012}. The main objective of the work is to demonstrate the capabilities of the proposed specific SG on the basis of WS-PGRADE platform and gUSE technology for MD simulations and further data post-processing on the basis of LAMMPS package for molecular dynamics (MD) simulation \cite{lammps1995}, and other packages like R for statistical analysis \cite{R2008}, Pizza.py Toolkit for manipulations with atomic coordinates files \cite{pizza2005}, AtomEye for visualisation of atomic configurations \cite{AtomEye2003}, debyer for simulation of X-ray diffraction (XRD) and neutron diffraction (ND) analysis (https://code.google.com/p/debyer/), etc. Several typical workflows were created for simulation of several physical processes with various demands for the computing resources: tension of metal nanocrystals under different physical conditions, tension of ensemble of metal nanocrystals under the same conditions, manipulations with complex nanostructures like indentation of graphene membranes, thermal stability of BNNT, etc.

\section{Background and Related Works}

The user communities outside of computer science itself (for example, in materials science, physics, chemistry, biology, nanotechnologies, etc.) would like to access various DCIs (grids, clouds, clusters) in a such transparent way, that they do not need to learn the peculiarities of these DCIs. Usually, various web-interfaces and portals are used for these purposes, for example, MolDynGrid \cite{MolDynGrid2009} or, especially, SCMS.pro (an open-source project) \cite{CMS2010}, which allow users to perform effectively many management operations with jobs, resources, users, etc.  The modern tendency is that scientists would want to concentrate on their scientific applications and get not only convenient management tools (like in the aforementioned solutions MolDynGrid and SCMS.pro), but also some high-level development frameworks for construction and management of complex scientific workflows with a tunable set of management tools in a shape of independent entities (like portlets). In this context, ``science gateway'' (SG) ideology means a user-friendly intuitive interface between scientists (or scientific communities) and various DCIs, where researchers can focus on their scientific goals and less on peculiarities of DCI. Usually, SGs are not specialized for a certain scientific area and scientists from many different fields can use them. The most important goals of SG are:
\begin{itemize}
  \item to provide a simplified intuitive graphical user interface (GUI) that is highly tailored to the needs of the given scientific community,
  \item to provide a smooth access to national and international computing and storage resources,
  \item to combine, create, and use collaborative tools for sharing scientific data on international scale.
\end{itemize}

\begin{figure}[hbtp]
  \begin{center}
    \includegraphics[width=0.5\textwidth]{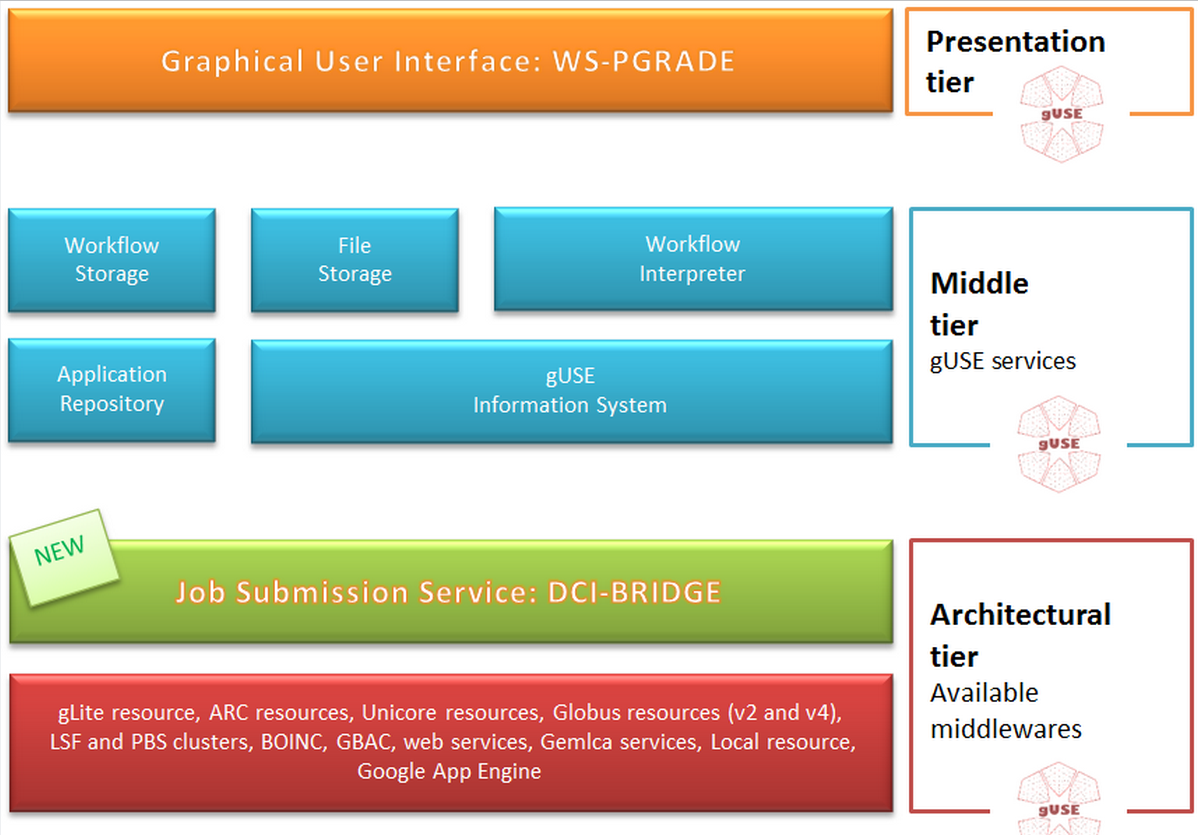}
    \caption{Architecture of science gateway ideology on the basis of gUSE+WS-PGRADE+DCI-Bridge technologies (https://guse.sztaki.hu) \cite{wspgrade2012}.}
    \label{fig:Portal_Architecture}
  \end{center}
\end{figure}

Several popular SG technologies are known for development of SG frameworks and instances, for example, WS-PGRADE/gUSE/DCI-bridge \cite{wspgrade2012}, ASKALON \cite{askalon2005}, MOTEUR \cite{moteur2008}, etc. They use different enabling components and technologies: web application containers (Tomcat, Glassfish, etc.), portal or web application frameworks (Liferay, Spring, Drupal, etc.), database management systems (MySQL, etc.), workflow management systems \cite{wfreview:deelman2009}. Here the most promising SG framework, namely WS-PGRADE/gUSE/DCI-bridge bundle (which is an open-source project), is described with emphasize on some examples of its usage in several scientific communities like physics, materials science, nanotechnologies (Fig.~\ref{fig:Portal_Architecture}).

WS-PGRADE/gUSE/DC-bridge framework \cite{wspgrade2012} represents the complex system with many advanced features (Fig.~\ref{fig:Portal_Architecture}). WS-PGRADE portal technology is a web based front end of the gUSE (grid and cloud User Support Environment), which is a well-known and permanently improving open source SG framework developed by Laboratory of Parallel and Distributed Systems (MTA-SZTAKI, Budapest, Hungary), that provides the convenient and easy access to DCIs. It gives the generic purpose, workflow-oriented GUI to develop and run workflows on various DCIs. WS-PGRADE supports development and submission of distributed applications executed on the computational resources of DCIs. The resources of DCIs (including clusters, service grids, desktop grids, clouds) are connected to the gUSE by a single point back end, the DCI-Bridge \cite{wspgrade2012}. The main idea of WS-PGRADE/gUSE/DCI-bridge approach is: ``The Portal is within reach of anyone from anywhere'' \cite{wspgrade:manual2013}.

\section{Science Gateway Internals}
\begin{figure}[hbtp]
  \begin{center}
    \includegraphics[width=0.5\textwidth]{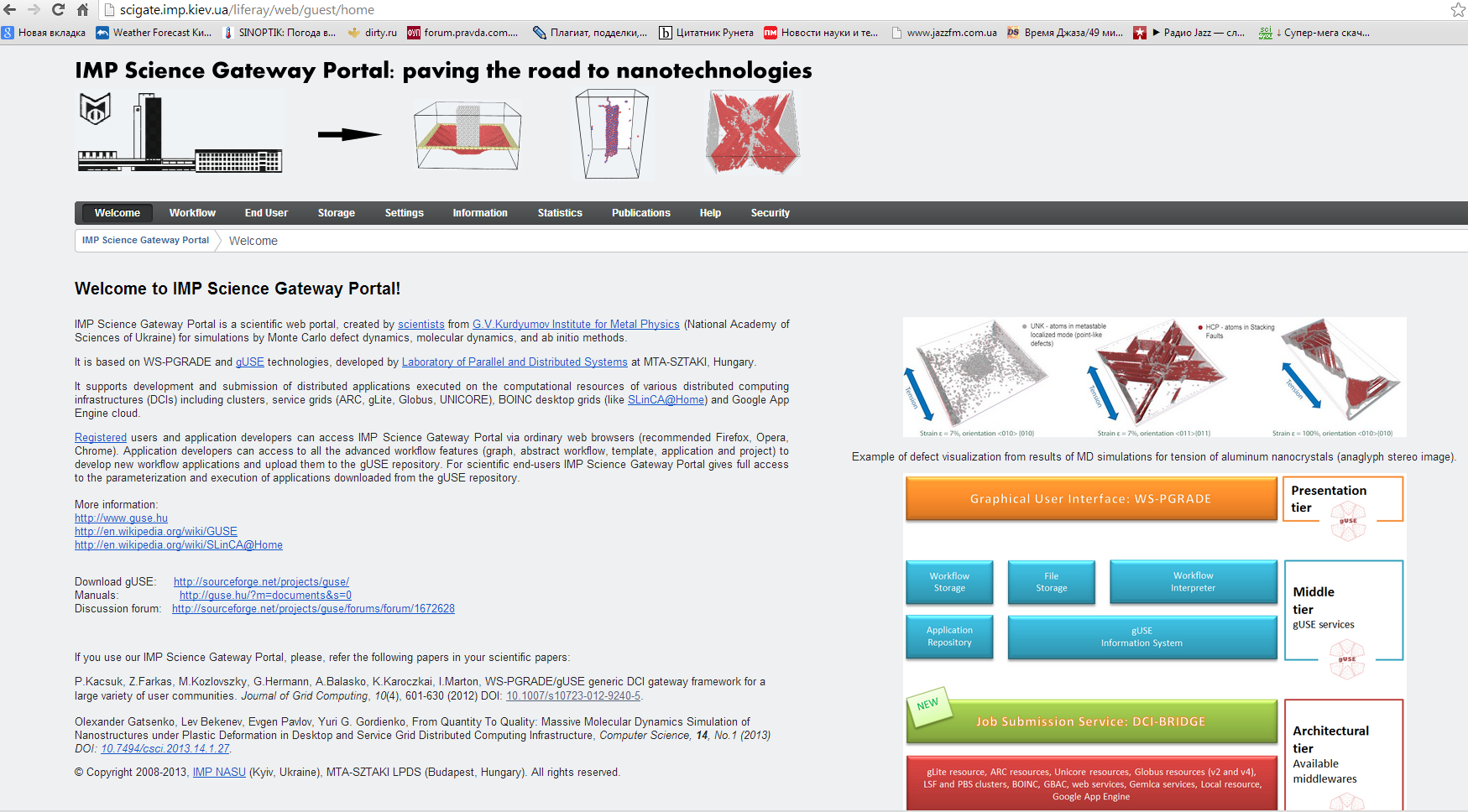}
    \caption{The graphical user interface of ``IMP Science Gateway Portal''.}
    \label{fig:Portal_GUI}
  \end{center}
\end{figure}

``IMP Science Gateway Portal'' (Fig. \ref{fig:Portal_GUI}) at the premises of G.V.Kurdyumov Institute for Metal Physics (IMP) National Academy of Sciences of Ukraine, is based on the WS-PGRADE/gUSE/DCI-bridge framework. The application developers can obtain access to the advanced workflow development features (graph, abstract workflow, template, application and project) to develop new workflow applications. The built-in repository stores the workflow and their components published by the application developers, which can be shared among user communities. But there are no ready workflows or their templates for a certain scientific area (like physics, materials science, or nanotechnologies) in the standard bundle of WS-PGRADE/gUSE/DC-bridge. That is why we designed, developed, tested, managed, and used several scientific workflows for different scientific applications (described below), and described in details the very complex workflow for MD simulations with complicated post-processing with many independent components for post-processing and data visualization (with various requirements, like input/output data format, parameters of running environment, etc.).

\begin{figure}[hbtp]
  \begin{center}
    \includegraphics[width=0.5\textwidth]{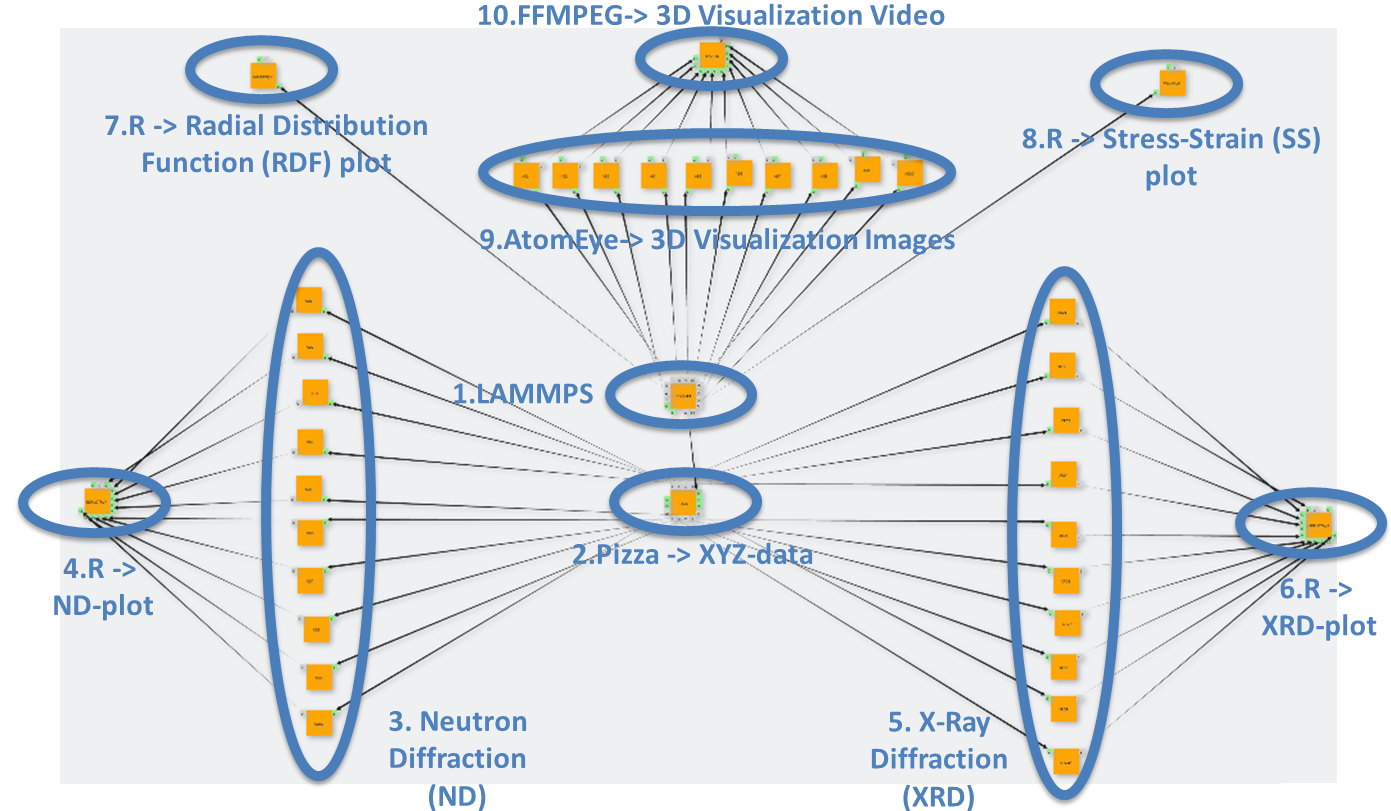}
    \caption{The complex workflow with executables (yellow bricks) and dataflows (arrows) created by means of WS-PGRADE tools for MD simulations.}
    \label{fig:Portal_WF_scheme}
  \end{center}
\end{figure}

\paragraph{Workflow Management.} ``IMP Science Gateway Portal'' provides easy and simple GUI utilities (on the basis of WS-PGRADE portlet) for application developers and end users to create workflow components (graph, abstract workflow, template, application and project) (Fig.~\ref{fig:Portal_WF_scheme}).
\begin{figure}[hbtp]
  \begin{center}
    \includegraphics[width=0.5\textwidth]{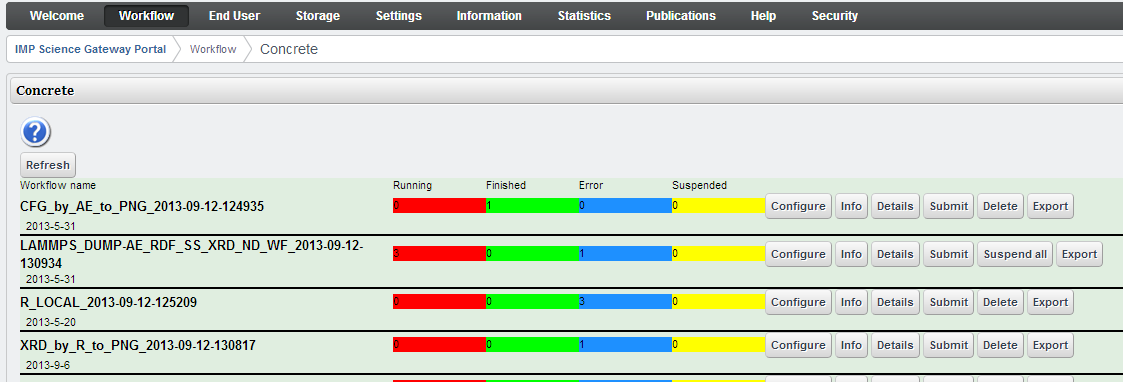}         a) \\
    \includegraphics[width=0.5\textwidth]{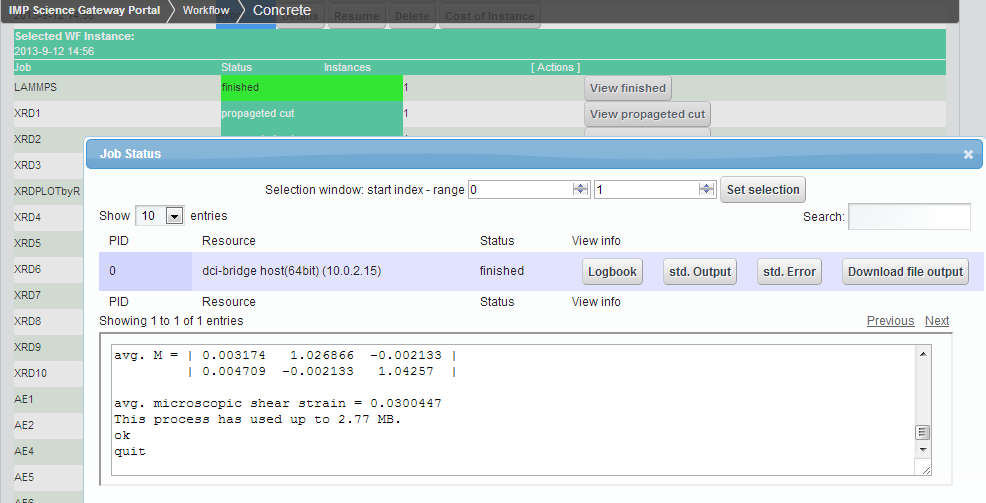}         b)
    \caption{The scientific workflow management in ``IMP Science Gateway Portal'' on the basis of WS-PGRADE portlet:  workflow list with options of operating (a) and monitoring abilities (b).}
    \label{fig:Portal_WF_management}
  \end{center}
\end{figure}

\begin{figure}[ht]
  \begin{center}
    \includegraphics[width=0.8\textwidth]{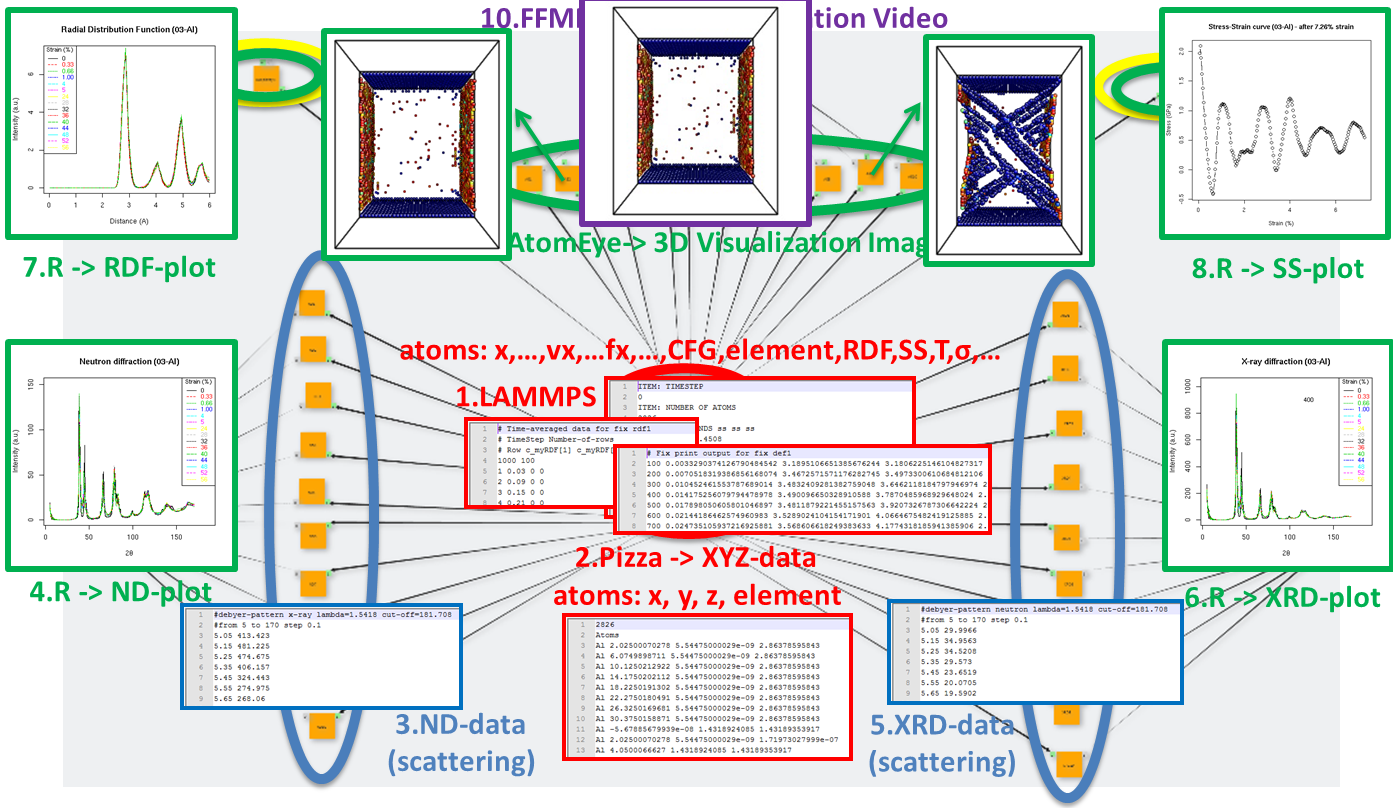}
    \caption{The workflow from Fig.~\ref{fig:Portal_WF_scheme} with some examples of input/output data.}
    \label{fig:AL_WF_scheme}
  \end{center}
\end{figure}

It provides for application developers and end users workflow management options to create, submit, monitor, save, upload, and retrieve various workflows with one or many jobs (Fig.~\ref{fig:Portal_WF_management}).
The process of the complex workflow creation (Fig.~\ref{fig:Portal_WF_scheme}) and management (Fig.~\ref{fig:Portal_WF_management}) can be demonstrated by the MD example of tension of metal nanocrystals with the following steps:

\begin{itemize}
  \item registration and signing in ``IMP Science Gateway Portal'';
  \item create a graph scheme for the basic blocks of the workflow (yellow bricks with small squares in Fig.~\ref{fig:Portal_WF_scheme});
  \item determine (create and name) a workflow from the graph;
  \item configure the workflow (links between yellow bricks in Fig.~\ref{fig:Portal_WF_scheme});
  \item start execution of the workflow;
  \item monitor the current status of execution (Fig.~\ref{fig:Portal_WF_management});
  \item check the MD simulation results obtained by LAMMPS package (tables in the red rectangles in the center of Fig.~\ref{fig:AL_WF_scheme});
  \item check the MD simulations results converted by Pizza.py Toolkit script package for post-processing (table in the red rectangle in the bottom of Fig.~\ref{fig:AL_WF_scheme});
  \item check the intermediate post-processed visualizations of atomic positions obtained by AtomEye package (images in the top part of Fig.~\ref{fig:AL_WF_scheme});
  \item check the intermediate post-processed results of X-ray analysis (of the MD simulations results) obtained by debyer (tables in the blue rectangles in the bottom part of Fig.~\ref{fig:AL_WF_scheme});
  \item get the final post-processed results (of the MD simulations results) after X-ray analysis, radial distribution function (RDF) analysis, and stress-strain (SS) analysis obtained by R-package (plots in the green rectangles at the left and right sides of Fig.~\ref{fig:AL_WF_scheme});
  \item get the final post-processed video of visualizations of atomic positions obtained by FFMPEG package (video in the violet rectangle in the top part of Fig.~\ref{fig:AL_WF_scheme}).
\end{itemize}

The complicated workflow (Fig.~\ref{fig:AL_WF_scheme}) for this application contains several independent components (LAMMPS package, Pizza.py Toolkit, AtomEye package, debyer, R-package, FFMPEG), which can be combined in various sets for the purposes of the end users \cite{cgw2011,Gatsenko2013}.

\paragraph{Integration of Computing Resources.} gUSE allows to scientific communities to compose and execute their scientific application on various DCIs. Supported types of DCIs are clusters (PBS, LSF), cluster grids (ARC, gLite, GT2, GT4, GT5, UNICORE), supercomputers (e.g. via UNICORE), desktop grids (BOINC) and clouds (via CloudBroker Platform and GAE). At the moment ``IMP Science Gateway Portal'' use clusters (PBS), cluster grids (ARC), desktop grids (BOINC), and plan to use supercomputers and clouds.

\paragraph{User Management.} The development and execution functionalities are separated and fitted to the different expectations of the following two main user groups:
\begin{itemize}
  \item The ordinary scientists without extensive knowledge of computer science with limited power and permissions, like common scientists from materials science, physics, chemistry, etc. (mentioned as ``end users''). They need only restricted manipulation possibilities and obtain the scientific applications in the ready state to configure and submit them with minimal efforts.
  \item The designers of the scientific processes and related simulation workflows (referenced as ``developers'' or ``power users'') need to build and to fit the application to be as comfortable as possible for the end users.
\end{itemize}

\section{Typical Examples of Practical Applications}
\begin{figure}[ht]
  \begin{center}
    \includegraphics[height=0.3\textwidth]{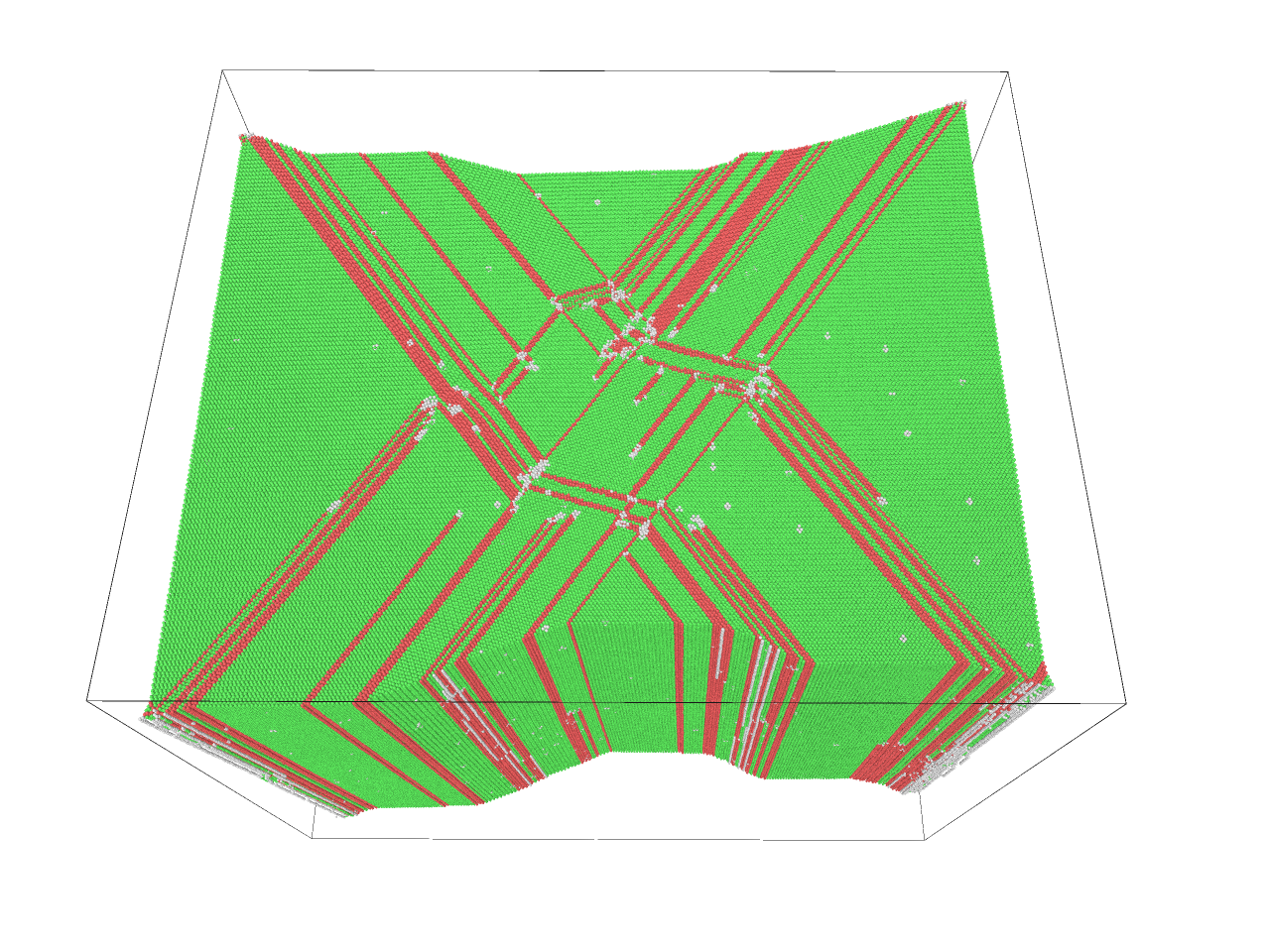} a)
    \includegraphics[height=0.3\textwidth]{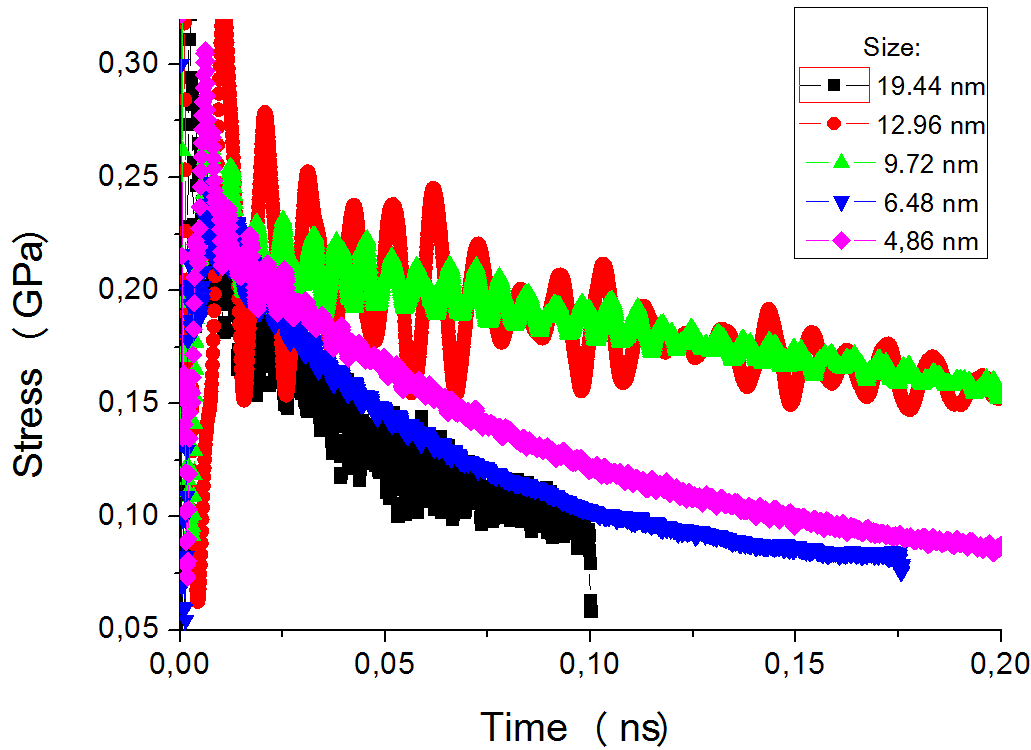} b)
    \caption{The evolution of the defect substructure in Al nanocrystal (after 15 ps, atoms in ideal lattice --- green color, stacking faults --- red color) (a), and oscillation of the internal stress $P_{xx}$ for different nanocrystal sizes (b).}
    \label{fig:al_oscillations}
  \end{center}
\end{figure}

Several practical applications in materials science are presented below where the different workflows were designed by WS-PGRADE workflow manager and used for computations in DCI by gUSE technology.
\paragraph{Mechanical Properties of Metal Nanocrystals.} This use case is related to MD simulation of relaxation behavior of nanocrystals after uniaxial tension (Fig.~\ref{fig:al_oscillations}a) \cite{gordienko2011,dubna2012}. The periodic time dependence of internal stress (Fig.~\ref{fig:al_oscillations}b) and the distribution of defects in the crystal structure was observed for different relaxing objects (single crystals of Al, Cu, and Si), physical conditions (tensile rate, size and orientation of the nanocrystals) and methodological parameters (potential type, boundary conditions, and others). The oscillatory nature of the relaxation of internal stresses is associated with the periodic rearrangement of the metastable defect substructure (point defects, dislocations, stacking faults and their intersections) (Fig.~\ref{fig:al_oscillations}b). These results confirm and significantly extend the known experimental data on the oscillatory nature of the relaxation processes and mechanisms of evolution of the initial defect substructure of crystals after ultrasound and magnetic effects \cite{nansys2013:steblenko}.

\begin{figure}[ht]
  \begin{center}
    \includegraphics[height=0.2\textwidth]{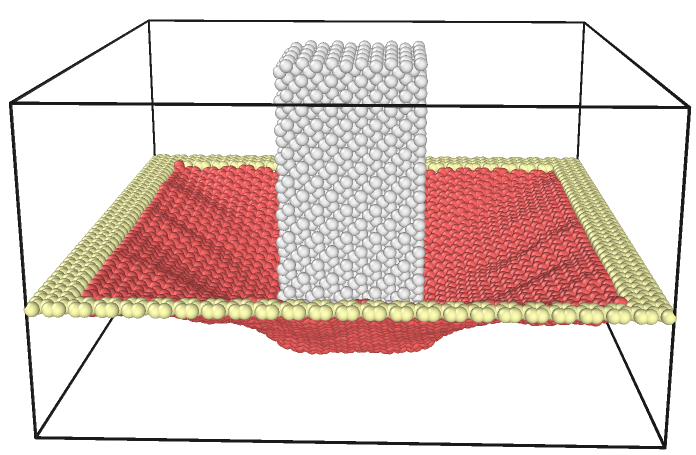} a)
    \includegraphics[height=0.3\textwidth]{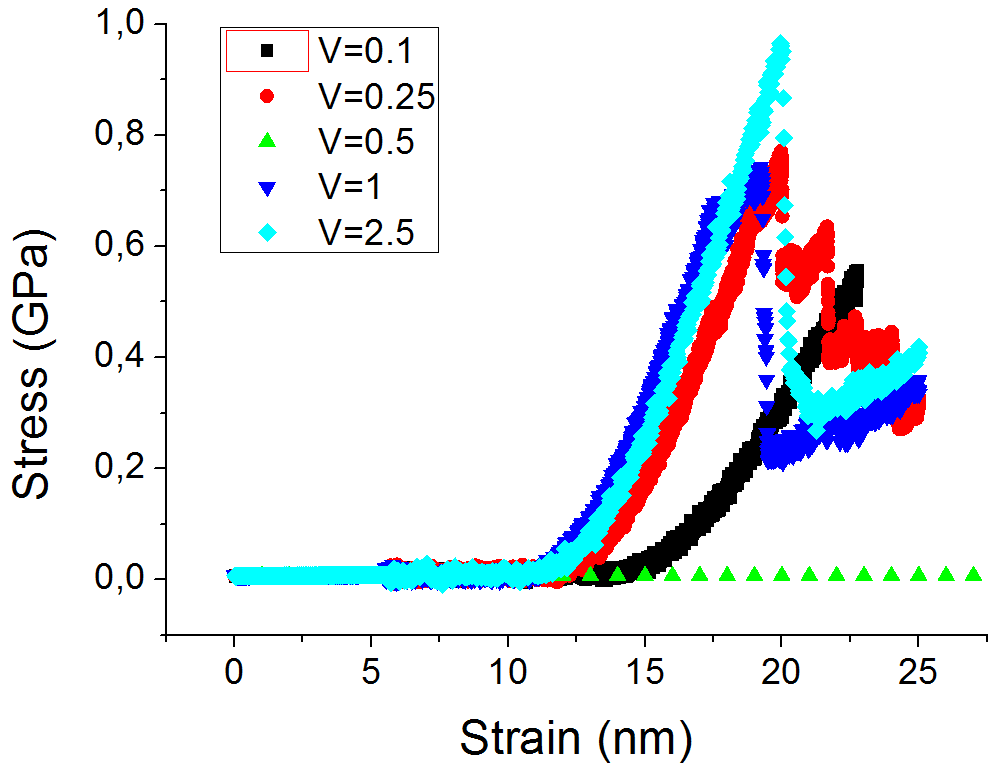} b)
    \caption{Nanoindentation of the graphene monolayer (mobile carbon atoms --- red color, fixed carbon atoms --- yellow color, nanoindentor atoms --- gray color) (a) for different speeds of the nanoindentor (b).}
    \label{fig:nanoindenation}
  \end{center}
\end{figure}

\paragraph{Elastic Properties of Graphene.} MD simulation of nanoindentation was performed for monolayer graphene membrane in an atomic force microscope (Fig.~\ref{fig:nanoindenation}a). It is shown that graphene membrane deforms and collapses within a wide range of strain values and applied forces (within the range 1.5-3 mN); internal tensile stresses along the axis - 0.6-0.8 GPa (Fig.~\ref{fig:nanoindenation}b). For a defect-free graphene under uniaxial tension, this corresponds to the tensile strength of 80-150 GPa and Young's modulus of 0.8-1.2 TPa. These results confirm and significantly extend the known experimental data of nanoindentation in a narrow range of parameters and the previous results of computer simulation of uniaxial tension of defect-free graphene \cite{nansys2013:tatarenko}.

\begin{figure}[ht]
  \begin{center}
    \includegraphics[height=0.3\textwidth]{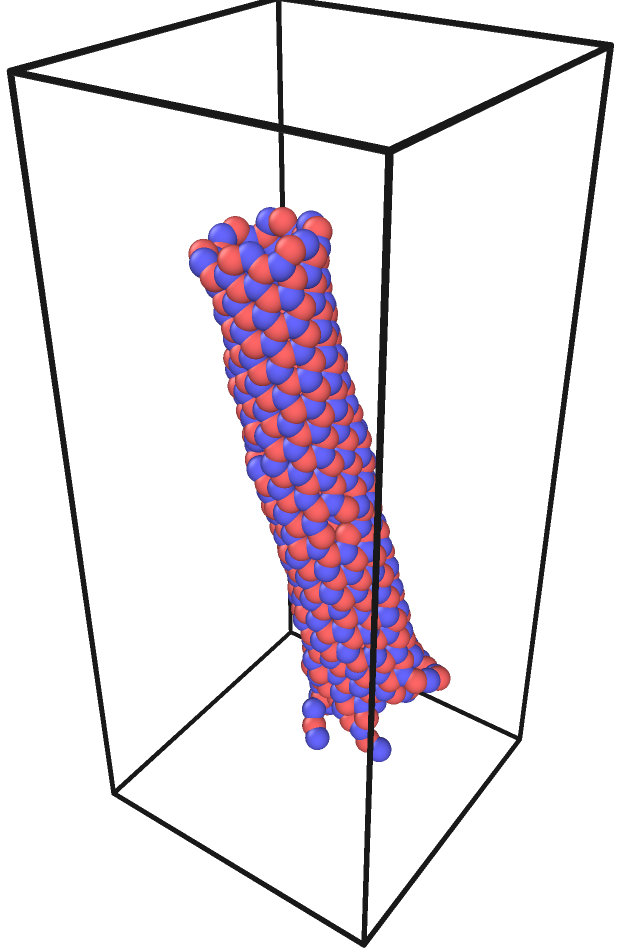} a)
    \includegraphics[height=0.3\textwidth]{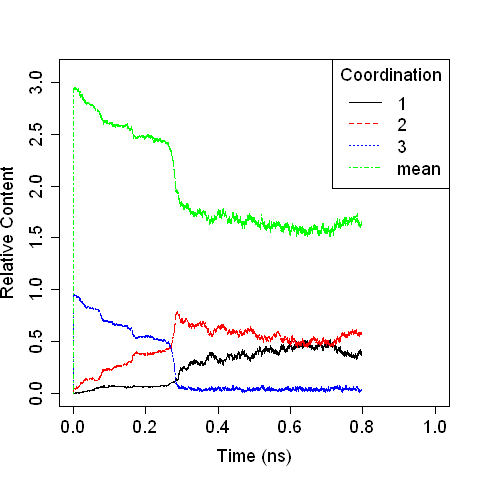} b)
    \caption{The collapse of BNNT (nitrogen atoms --- red color, boron atoms --- blue color) (a), the dynamics of the collapse for BNNT characterized by atom coordination numbers (b).}
    \label{fig:BNNT}
  \end{center}
\end{figure}

\paragraph{Thermal Properties of Boron Nitride Nanotubes.} Boron nitride nanotubes (BNNT) possess exceptional physical properties, which are a prerequisite for their wide practical applications in the future (Fig.~\ref{fig:BNNT}a). MD simulation workflow was used for investigation of thermal stability of BNNT with different chirality, size and perfection of structure (Fig.~\ref{fig:BNNT}a). The dynamics (Fig.~\ref{fig:BNNT}b) and temperature dependence of the BNNT collapse was observed for BNNTs with various chiralities, sizes and perfection of structure, as well as the simulation parameters (type of potential, boundary conditions, thermal conditions, etc.). It is shown that the BNNT collapse proceeds by separation of atoms from BNNT open caps and BNNT collapse depends on their chirality (zigzag, armchair, etc.), but the mechanism of their collapse (by separation of atoms from the open caps) remains unchanged. These results confirm and significantly extend the known experimental data on the thermal stability of BNNTs \cite{nansys2013:sartinska}.

\section{Conclusion}
The general purpose and application specific workflows for computer simulations of complex natural processes in materials science, physics, and nanotechnologies can be easily designed and used in practice on the basis of the available ``science gateway'' technologies. One of them was demonstrated by ``IMP Science Gateway Portal'' created on the basis of: WS-PGRADE technology for workflow management and gUSE technology for parallel and distributed computing in various heterogeneous DCIs. Several typical scientific applications were considered as use cases of its usage. Porting MD-applications to heterogeneous DCI by means of ``science gateway'' ideology is easy and efficient, if WS-PGRADE platform is used, and parameter decomposition and sweeping parallelism are possible. As a result, MD simulations of complex behavior of various nanostructures (graphene, metal nanocrystals, boron nitride nanotubes, etc.) can be effectively carried out in the heterogeneous DCI with a low efforts and quite short learning curve.

The obvious advantages are as follows: smooth access to heterogeneous DCI and software, division of user roles (administrators: operators of portals; power users: principal scientists; end users: scientists, students), possibility to make more complex workflows with intuitive GUI (with additional modules, ad hoc changes, etc.), short learning curve for usual scientists without extensive knowledge in computer science.

Some non-critical drawbacks of the approach include: non-standard file naming convention (alphanumeric only), which can cause problems with legacy code with special symbols, tacit "stdout" and "stderr" information style for some errors in WS-PGRADE, tacit style of manuals and context help. All of them can be alleviated by the proper improvement of some non-critical components and general documentation.

The ``science gateway'' approach -- workflow manager (like WS-PGRADE) + DCI resources manager (like gUSE) at the premises of the portal (like ``IMP Science Gateway Portal'') -- is very promising in the context of its practical MD applications in materials science, physics, chemistry, biology, and nanotechnologies.

\section{Acknowledgments}
The work presented here was partially funded by EU FP7 SCI-BUS (SCIentific gateway Based User Support) project, No. RI-283481, and partially supported in the framework of the research theme "Introduction and Use of Grid Technology in Scientific Research of IMP NASU" under the State Targeted Scientific and Technical Program to Implement Grid Technology in 2009-2013.

\end{document}